\documentclass[11pt,a4paper]{article}
\pdfoutput=1
\usepackage{jheppub}

\usepackage{amsmath}
\usepackage{amssymb}
\usepackage{amsthm}
\usepackage{MnSymbol}
\usepackage{nicefrac}
\usepackage{amsfonts}
\usepackage{dsfont}

\usepackage[markup=nocolor]{changes}
\usepackage{natbib}

\usepackage{graphicx}
\graphicspath{{Figures/}}  

\usepackage[normalem]{ulem}

\usepackage{dcolumn}
\usepackage{bm}
\usepackage{color}
\usepackage{slashed}
\usepackage{empheq}
\newcommand{\bea}{\begin{eqnarray}}
\newcommand{\eea}{\end{eqnarray}}
\newcommand{\be}{\begin{equation}}
\newcommand{\ee}{\end{equation}}

\title{
Predictive power of grand unification from quantum gravity
}

 \author[a,b]{Astrid Eichhorn}
   \emailAdd{eichhorn@sdu.dk}
\affiliation[a]{
CP3-Origins, University of Southern Denmark,\\Campusvej 55, DK-5230 Odense M, Denmark}
\affiliation[b]{Institut f\"ur Theoretische
  Physik, Universit\"at Heidelberg,\\Philosophenweg 16, 69120
  Heidelberg, Germany}
\affiliation[c]{Theoretical Physics, Blackett Laboratory,
Imperial College,\\London, SW7 2AZ, U.K.}
\author[b,c]{Aaron Held}
\emailAdd{a.held@imperial.ac.uk}
  \author[b]{Christof Wetterich}
\emailAdd{c.wetterich@thphys.uni-heidelberg.de}

\abstract{
If a grand-unified extension of the asymptotically safe Reuter fixed-point for quantum gravity exists, it determines free parameters of the grand-unified scalar potential. All quartic couplings take their fixed-point values in the trans-Planckian regime. They are irrelevant parameters that are, in principle, computable for a given particle content of the grand unified model. In turn, the direction of spontaneous breaking of the grand-unified gauge symmetry becomes predictable. For the flow of the couplings below the Planck mass, gauge and Yukawa interactions compete for the determination of the minimum of the effective potential.
}

\begin{document}
\maketitle

\section{Introduction}
A grand unified theory (GUT) unifies the gauge group $SU(3)\otimes SU(2) \otimes U(1)$ of the Standard Model (SM) into a larger simple group \cite{Georgi:1974sy, Pati:1974yy, Fritzsch:1974nn}, typically $SU(5)$, $SO(10)$ or $E_6$. This leads to the prediction of (almost) equal gauge couplings at the scale $M_X$, at which the GUT-group is spontaneously broken. One, therefore, obtains two relations between the strong, weak, and electromagnetic gauge couplings in dependence on $M_X$. From the observed values of these couplings, one infers that the ratio $M_X/M_W$, with $M_W$ the $W$-boson mass, needs to be huge (gauge hierarchy), explaining why the predicted decay of the proton is small enough to not have been observed until today. 
For $SO(10)$ all chiral fermions of the SM of a given generation belong to the same $16$-dimensional representation, which includes an additional singlet neutrino. For $SU(5)$ the down-type quarks and the charged leptons belong to the same $5$-dimensional representation. For the third generation, a single scalar field may dominate the effective Yukawa couplings in the SM. In this case, GUTs  predict the ratio between bottom-quark and tau-lepton mass \cite{Buras:1977yy, Georgi:1979df, Georgi:1979ga, Lazarides:1980nt}.
\\

Despite the conceptual simplicity, the predictions for gauge couplings and partly Yukawa couplings, and insights on the structure of neutrino masses, further observational tests of the GUT-scenario have so far not been possible. This is a consequence of the GUT scale being out of reach of direct experimental tests. On the other hand, the GUT paradigm can only be ruled out by data at scales much below the GUT scale if the former has a sufficiently high predictive power. 
Yet, the standard GUT paradigm suffers from a notable absence of predictive power, as seen, e.g., in the many possible breaking chains connecting the SO(10) GUT to the SM. At the root of this predictivity problem are the many unknown couplings in the scalar sector of GUTs. Scalars are needed for a realistic chain of spontaneous symmetry breaking (SSB), see, e.g., \cite{Gipson:1984aj, Chang:1984qr, Deshpande:1992au, Deshpande:1992em, Bertolini:2009qj} for non-supersymmetric SO(10) breaking chains. Their masses, quartic couplings (and sometimes cubic couplings) are unknown renormalizable couplings. As a direct consequence, neither the scale $M_X$, nor the direction of GUT-symmetry breaking is determined uniquely. All group theoretically admissible breaking chains can be constructed and the magnitude of proton decay cannot be predicted.\\

The situation could change profoundly if one attempts to combine the GUT and gravity into a common renormalizable quantum field theory. All quartic scalar couplings, and perhaps also cubic couplings and mass terms, become predictable.
This requires the presence of an ultraviolet (UV) fixed point for which some interactions do not vanish, and at which the quartic couplings are irrelevant parameters. 
The asymptotic-safety scenario \cite{Weinberg:1980gg} has become a much-explored possibility in four Euclidean dimensions due to the pioneering work of Reuter \cite{Reuter:1996cp}, based on functional renormalization of the effective average action \cite{Wetterich:1992yh, Ellwanger:1993mw, Morris:1993qb,Reuter:1993kw,Tetradis:1993ts}. The UV fixed point, referred to as the Reuter fixed point, is seen in systematic approximations both for pure gravity \cite{Reuter:2001ag, Lauscher:2001ya, Litim:2003vp, Codello:2008vh, Benedetti:2009rx, Manrique:2010am, Manrique:2011jc, Falls:2013bv, Christiansen:2014raa,Falls:2014tra, Becker:2014qya, Gies:2015tca, Christiansen:2015rva,Demmel:2015oqa, Ohta:2015efa,Gies:2016con, Biemans:2016rvp,Denz:2016qks, Christiansen:2017bsy, Falls:2017lst, Gonzalez-Martin:2017gza,Falls:2018ylp,deBrito:2018jxt,Knorr:2019atm,Bosma:2019aiu} 
and for gravity coupled to matter \cite{Narain:2009fy, Daum:2009dn,Harst:2011zx, Dona:2013qba, Dona:2015tnf, Oda:2015sma, Percacci:2015wwa,  Meibohm:2015twa, Labus:2015ska, Eichhorn:2016vvy, Biemans:2017zca,Hamada:2017rvn, Christiansen:2017cxa, Eichhorn:2018ydy, Eichhorn:2018nda, Alkofer:2018fxj,Pawlowski:2018ixd,Wetterich:2019zdo}. 
 While the functional Renormalization Group (FRG) is the most commonly used tool to explore asymptotically safe gravity, it is by no means the only one. Lattice techniques \cite{Loll:2019rdj}, also with matter \cite{Jha:2018xjh,Catterall:2018dns} and perturbative tools \cite{Niedermaier:2009zz,Niedermaier:2010zz} are also employed. \\
 Within the FRG approach to asymptotic safety considerable progress has been achieved within the last years, in particular, in the development of extended truncations, following the guiding principle of near-canonical scaling \cite{Falls:2013bv,Falls:2014tra,Falls:2017lst,Falls:2018ylp,Eichhorn:2018akn,Eichhorn:2018ydy,Eichhorn:2018nda}.
This is closely analogous to similar developments, e.g., in the Ising universality class, where FRG techniques give rise to quantitatively precise results \cite{Balog:2019rrg}.
This has given rise to compelling evidence for the existence of the Reuter fixed point in four-dimensional Euclidean gravity. 
A number of important open questions exist, which include the fate of the fixed point in Lorentzian signature, its compatibility with background-independence and the presence of potential tachyonic- or ghost-like instabilities, see \cite{Bonanno:2020bil} for a comprehensive critical appraisal of the state of the art in asymptotically safe quantum gravity, and \cite{Donoghue:2019clr} in particular for a discussion of Lorentzian signature. Most importantly, besides questions of internal consistency, a key challenge for quantum gravity is its consistency with observable phenomenology.
It is this latter challenge that we address in the present paper. \\
  While it is probably true that experimental ``smoking-gun" signatures of quantum gravity are difficult to achieve (see \cite{Bonanno:2017pkg,Wetterich:2019qzx,Rubio:2017gty,Platania:2020lqb} for recent accounts of potential effects in cosmology), consistency tests with low-energy observations could actually provide a critical test of quantum gravity. In particular, asymptotically safe quantum gravity might be rather restrictive with regard to the permissible properties of matter sectors.
Potential phenomenological implications for the matter sector have been explored, e.g., in \cite{Shaposhnikov:2009pv,Harst:2011zx,Eichhorn:2017lry,Eichhorn:2017ylw,Eichhorn:2017als,Eichhorn:2018whv,Bonanno:2018gck,Grabowski:2018fjj,deBrito:2019epw,Kwapisz:2019wrl}.
Due to non-vanishing interactions at the fixed point, the simple association of relevant couplings with perturbatively renormalizable couplings, having zero or positive mass dimension, is no longer possible. Some renormalizable couplings of the SM or GUT may correspond to irrelevant parameters at the UV fixed point and therefore become predictable \cite{Shaposhnikov:2009pv,Harst:2011zx,Wetterich:2016uxm,Eichhorn:2017ylw,Eichhorn:2017lry,Eichhorn:2017egq,Eichhorn:2017muy,Eichhorn:2018whv}. An intuitive way to understand the enhanced predictive power is the consideration that enhanced symmetries in QFTs typically lead to restrictions on the values of allowed interactions. In the case of asymptotic safety, the symmetry is quantum scale symmetry \cite{Wetterich:2019qzx} in the UV. 
For a recent review of asymptotic safety in QFTs, see \cite{Eichhorn:2018yfc}; for reviews of asymptotically safe gravity (with matter) see, e.g., \cite{Niedermaier:2006ns,Reuter:2012id,Percacci:2017fkn,Eichhorn:2017egq,Pereira:2019dbn,Reuter:2019byg}.
\\

The recent renewed interest in GUTs, see, e.g., \cite{Klein:2019jgb}, and \cite{Croon:2019kpe} for a recent review,  includes the exploration of asymptotic safety in GUTs without gravity \cite{Bajc:2016efj,Molinaro:2018kjz,Wang:2018yer,Abel:2017rwl} as well as the coupling of asymototically free GUTs \cite{Cheng:1973nv} to quadratic gravity \cite{Giudice:2014tma, Holdom:2014hla, Einhorn:2016mws, Einhorn:2017jbs,Einhorn:2019dwm}. Throughout this paper, we will focus on the non-supersymmetric case whenever we discuss specific examples. For discussions of the phenomenological viability of minimal non-supersymmetric SO(10) models see, e.g., \cite{Harvey:1981hk, Babu:1992ia, Matsuda:2000zp, Matsuda:2001bg, Bajc:2005zf, Bertolini:2009es, Altarelli:2010at, Joshipura:2011nn, Bertolini:2012im, Altarelli:2013aqa, Dueck:2013gca, Babu:2015bna, Babu:2016bmy, Fukuyama:2018aub, Ohlsson:2018qpt, Schwichtenberg:2018cka, Ohlsson:2019sja}. If an extension of the Reuter fixed point to the supersymmetric GUT setting should exist, see \cite{Dona:2014pla}, our discussion would carry over to that case.

\section{Predictivity for quartic scalar couplings}
\label{sec:predictivity}

Asymptotically safe quantum gravity allows to predict
dimensionless quartic scalar couplings, as was first realized for the successful prediction \cite{Shaposhnikov:2009pv} of the Higgs-boson mass in the SM coupled to quantum gravity to be 126 GeV with an estimated few GeV uncertainty. This prediction assumes that the fixed-point value of the non-minimal coupling of the Higgs to the curvature scalar is sufficiently small.
The mechanism underlying this prediction is a gravity-induced anomalous dimension $A$, which contributes to the running of the quartic coupling $\lambda_H$ at transplanckian momentum scales $k>M_P$, where $M_P$ is the IR-value of the Planck mass, i.e., 
\begin{align}
    \label{eq:chr-A}
   \beta_{\lambda_H}= \partial_t \lambda_H = (A+ 2\eta)\lambda_H + \mathcal{C}_g - \mathcal{C}_y + c_\lambda\,\lambda_H^2\;.
\end{align}
Here, $t=\ln(k/\bar{\mu})$ involves the infrared (IR) cutoff $k$ of functional renormalization which acts as a renormalization scale, and $\bar{\mu}$ is some arbitrarily fixed scale. The terms $\mathcal{C}_g$, $\mathcal{C}_y$ and $c_\lambda\,\lambda_H^2$ represent contributions from fluctuations of gauge bosons with gauge coupling $g$, fermions with a Yukawa coupling $y$, and scalars. To leading order, $\mathcal{C}_g$ depends on $\alpha_g = g^2/(4\pi)$ quadratically, i.e., $\mathcal{C}_g\sim\alpha_g^2$, while $\mathcal{C}_y$ is proportional to $\alpha_y^2$, with $\alpha_y = y^2/(4\pi)$, and $c_\lambda$ is a positive constant. 
Assuming a fixed point at which gauge- and Yukawa-couplings vanish, both $\mathcal{C}_g$ and $\mathcal{C}_y$ vanish in the UV and the only available fixed point for $\lambda_H$ lies at zero quartic coupling, $\lambda_H = 0$. This is in line with symmetry considerations: There are no indications that asymptotically safe gravity breaks global symmetries in the matter sector. Thus, a fixed point that preserves shift-symmetry for the scalar, and thus corresponds to a flat scalar potential, is expected to exist.
For $\alpha_g = \alpha_y = 0$, also the gauge-boson and fermion-induced anomalous dimension $\eta$ vanishes. The gravity-induced anomalous dimension $A$ is positive \cite{Narain:2009fy, Percacci:2015wwa, Labus:2015ska, Oda:2015sma, Wetterich:2016uxm, Wetterich:2017ixo, Hamada:2017rvn, Eichhorn:2017als, Pawlowski:2018ixd, Wetterich:2019zdo} (see \cite{deBrito:2019umw} for restrictions on higher-order gravitational couplings arising from this requirement), quickly driving any non-zero transplanckian $\lambda_H$ towards small values as $k$ is lowered. This follows from the IR attractive nature of this fixed point.\\

Under the assumption that the fundamental microscopic description of quantum gravity is based on the UV fixed point, any irrelevant coupling approaches its fixed-point value during an infinite flow interval\footnote{Note that even in settings where asymptotic safety only holds approximately, i.e., the fundamental microscopic theory is a different one, but the flow in the effective description remains close to the fixed point for a range of scales, irrelevant couplings 
approach their fixed-point values, see \cite{deAlwis:2019aud} for a specific scenario.}. 
As a result, all irrelevant couplings are predicted to take their fixed-point values all the way down to $M_P$. For positive $A$ the quartic coupling $\lambda_H$ is an irrelevant coupling. 
At all transplanckian scales, not only $\lambda_H$ vanishes, but all other quartic and higher-order couplings vanish. The situation is slightly more subtle for mass-terms: Their fixed-point values vanish, but they may deviate from these fixed-point values already at transplanckian scales, if they are relevant couplings. Whether the gravitational contribution is large enough to counteract the canonical dimension and render the mass terms irrelevant depends on the exact position of the gravitational fixed point, see, e.g., \cite{Wetterich:2016uxm, Eichhorn:2017als, Wetterich:2019zdo}. Both possibilities -- a relevant mass term with reduced scaling dimension as well as an irrelevant one -- lie within the current systematic uncertainties of the location of the gravitational fixed point. Under the assumption of an irrelevant mass term, the whole scalar potential $U$ is flat for all $k\geq M_P$, given by a field-independent value $U(\rho_i) = u_\ast\,k^4$.
Here, the $\rho_i$ are invariants formed from scalar fields, such as $\rho = h^\dagger h$ for the Higgs doublet $h$. Under the assumption of a relevant mass term, a flat potential can also be achieved at the Planck scale. This is only one of many possibilities, as arbitrary finite values of the quadratic term at the Planck scale are compatible with asymptotic safety in this case. With an (almost) flat effective potential for $k\approx M_P$, determined by quantum-gravity fluctuations and matter fluctuations, the final shape of the effective potential at $k \approx 0$ will be determined by the matter fluctuations.
\\

The generalization to a fixed point with nonzero gauge and Yukawa interactions, $\alpha_{g\,\ast}\neq 0$ and/or $\alpha_{y\,\ast}\neq 0$, works as follows. In this case, the fixed-point values of $\mathcal{C}_g$ and $\mathcal{C}_y$ are constants. For small $\alpha_g$ and $\alpha_y$, they are proportional to $\alpha_{g^2\ast}$ and $\alpha_{y^2\ast}$, respectively. Further, $\eta$ also contains terms $\sim\alpha_g$ and $\sim\alpha_y$, but those are subleading compared to $\mathcal{C}_{g/y}$, such that we neglect them here. To leading order in small $\alpha_g,\,\alpha_y$, one finds a slightly shifted fixed point 
\begin{align}
    \label{eq:lambda-FP}
    \lambda_{H\, \ast} = \frac{\mathcal{C}_y - \mathcal{C}_g}{A}\;.
\end{align}
The deviation $\lambda_H - \lambda_{H\,\ast}$ is again an irrelevant coupling, with critical exponent $-A$, as can be seen by inspecting $-\partial_{\lambda_H}\beta_{\lambda_H}$ at the fixed point.
Accordingly, the quartic scalar coupling assumes the nonvanishing
fixed-point value given by Eq.~\eqref{eq:lambda-FP} at transplanckian scales, $k\geq M_P$.
The discussion generalizes to models with several scalar couplings $\lambda_i$. The gravity-induced anomalous dimension $A$ is universal (here meant to signify independent of matter couplings and internal symmetries), such that all quartic couplings are predicted to take fixed-point values approximately given by Eq.~\eqref{eq:lambda-FP}, now with potentially different $\mathcal{C}_g^{(i)}$ and $\mathcal{C}_y^{(i)}$ for the different couplings. In cases where the scalar couplings contribute to the beta functions among each other, Eq.~\eqref{eq:lambda-FP} receives additional contributions. For $A$ of the order one, $|\lambda_{H\ast}|$ is a very small quantity for realistic gauge and Yukawa couplings.\\

The main modification of the running of the couplings $\lambda_i$ as compared to a quantum field theory without gravity is the addition of the gravity-induced anomalous dimension, i.e., a term $A\,\lambda_i$  in the $\beta$-functions $\beta_i = \partial_t\,\lambda_i$. Let us highlight that $A$ is independent of the internal symmetries, which is simply a consequence of the ``blindness" of gravity to internal symmetries. Therefore, $A$ does not need to be recomputed if one changes the internal symmetry group and representation of the scalar. On the other hand, it is well-known that $A$ is non-universal in the sense of scheme-independence. Since the gravitational coupling is dimensionful, computations in massless schemes can be less complete.
For instance, \cite{Rodigast:2009zj} have computed gravitational corrections to the quartic scalar beta function to first order in perturbation theory, which were found to vanish, unless there is also a scalar mass present. Let us note that non-universality affects any beta function, also in non-gravitational settings. For dimensionful couplings, it sets in at leading order in perturbation theory, for dimensionless ones in a mass-independent RG scheme, it sets in at the next-to-next-to-leading order in perturbation theory. This applies, for instance to the Standard Model, as well as the non-gravitational part of the flow in GUTs. Given that neither beta functions nor the scale-dependent couplings themselves are directly observables (although they of course enter their computation), such a non-universality is neither unexpected nor problematic. At the level of observables, such dependences cancel. The cancellation is imperfect, as long as calculations are done in approximations. This is being used to estimate the quality of an approximation, see, e.g., \cite{Gies:2015tca,deBrito:2018jxt} for examples within the FRG for gravity and \cite{Balog:2019rrg} for a non-gravitational example. \\
In our context, it is key that critical exponents, unlike beta functions, are universal. Accordingly, the number of predictions that can be made in a given setting, encoded in the number of negative critical exponents, is independent of which scheme one chooses to compute in. Note that this does not require the beta functions themselves to agree; their form can be rather different in different schemes while the critical exponents remain the same.
In our case, this implies in particular that the number of additional predictions that can be made for scalar potentials once the gravitational contribution is included, is expected to be independent of the scheme one computes in. 
The FRG is expected to be a particularly well-suited scheme to compute gravitational corrections since it keeps track of mass-like thresholds in beta functions. Given that gravitational corrections come in with a typical mass scale, the Planck scale, this property of the FRG is expected to make it well-suited to compute those corrections. 

Ultimately, as in the paradigmatic examples of the Ising model \cite{Balog:2019rrg}, as well as the Gross-Neveu model in 3 dimensions \cite{Parisi:1975im,Gawedzki:1985ed,Gracey:1990sx,Hikami:1976at,Rosenstein:1988pt,Gat:1991bf,Rosenstein:1993zf,Mihaila:2017ble,Zerf:2017zqi,Rosa:2000ju,Hofling:2002hj,Braun:2010tt,Gies:2010st,Gehring:2015vja,Classen:2015mar,Vacca:2015nta,Knorr:2016sfs}, both of which feature interacting fixed points, with the latter being a main example for asymptotic safety, a convergence of results from different methods is expected for the critical exponents. In this context, it is promising that gravitational corrections to the scalar mass, computed with lattice techniques in Euclidean Dynamical Triangulations \cite{Jha:2018xjh}, agree with the FRG results that indicate that gravitational corrections on their own drive the scalar potential towards flatness, i.e., act in a shift-symmetry restoring way. Ultimately, similar studies in a Lorentzian approach to quantum gravity, such as causal set theory \cite{Bombelli:1987aa,Dowker:aza,Surya:2019ndm}, could also provide access to gravitational corrections for scalar potentials in the Lorentzian regime; see \cite{Glaser:2018jss} for first studies of the phase diagram of causal sets coupled to simple matter.

The functions $\mathcal{C}_g$, $\mathcal{C}_y$, $c_\lambda$, and $\eta$ may be computed in perturbation theory.
As we noted,
the dimensionful nature of the gravitational coupling makes gravitational contributions to beta functions non-universal. Therefore one should be careful not to mix schemes even at leading order. 
Our results are based on functional Renormalization Group techniques. 
The one-loop contributions arising from the dimensionless couplings in the matter theory without gravity 
are the same in any scheme.
Thus, for marginal couplings, functional Renormalization gives the same results, as, e.g., the MS-scheme at one loop.
Beyond this particular case, scheme matching is required for precise predictions.  Within functional renormalization, threshold functions allow for an easy description of the effective decoupling of particles with mass $m>k$.
As far as the issue of predictivity is concerned, the matching of scales is not needed.  We stress again that the critical exponents of a fixed point -- which encode the number of free parameters and thus predictions of the model -- are scheme-independent.\\

We observe that for slowly running $g$, $y$ and $\lambda$ and for slowly varying, large $A$, Eq.~\eqref{eq:lambda-FP} remains a very good approximation. The corresponding values $\lambda_i(k)$ now reflect approximate partial fixed points that only vary slowly with the scale $k$. The predictive power for the values of these couplings is kept if the model extends to $k\rightarrow\infty$. The addition of the gravity-induced anomalous dimensions to the flow equations is the basis for the predictivity of quantum gravity for all quartic couplings in GUT-models. This novel contribution is also the basis for the prediction of the quartic coupling of the Higgs in the SM  in \cite{Shaposhnikov:2009pv}.
Let us point out that the existence of a gravitational fixed point under the impact of quantum fluctuations of GUT matter requires further studies \cite{Dona:2013qba, Alkofer:2018fxj,Wetterich:2019zdo}. In this paper, we work under the assumption that a gravitational fixed point exists and explore its impact on the GUT. The requirement of a joint fixed point for GUT-gravity is expected to already result in restrictions on the allowed GUT matter content.
\\

\begin{figure}[!t]
\centering
\includegraphics[width=0.5\linewidth]{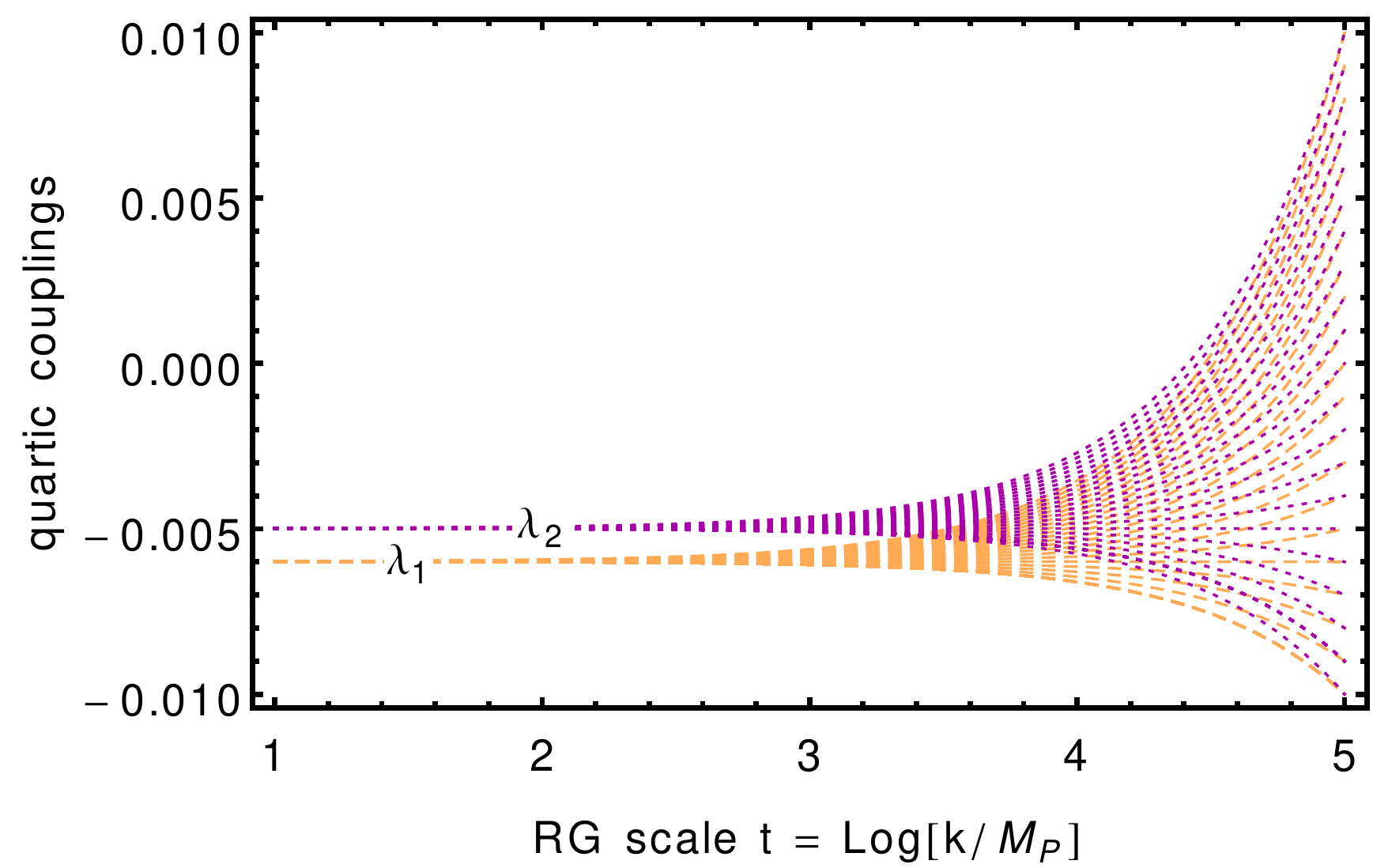}
\caption{\label{fig:adjointSU5_running}
Transplanckian running of the two quartic couplings of a scalar transforming in the adjoint  representation of SU(5), i.e., $\lambda_1$ and $\lambda_2$ as in Eq.~\eqref{eq:SUNpot}, as orange-dashed and purple-dotted lines, respectively. The running is obtained by numerically evolving Eqs.~\eqref{eq:SUNbetas-1},\eqref{eq:SUNbetas-2} with $N=5$, constant $A=2$ and constant $\alpha=1/40$. The quartic couplings have been initialized on a grid of points on the interval $[-0.01,0.01]$ each.
}
\end{figure}

As an example with multiple quartic couplings, we 
investigate the simplest 
GUT setup, namely $SU(N)$ unification with an $(N^2-1)$ dimensional scalar representation $\phi_\text{adj}$ responsible for the breaking. The latter can be represented by the set of $N\times N$ traceless and hermitian matrices. This gauge group could be relevant for beyond Standard-Model phenomenology in the case $N=5$.
The $24$-component adjoint representation can break $SU(5)$ to the SM-symmetry group $SU(3)\times SU(2)_L\times U(1)_Y$.
The quartic scalar potential is given by two invariants with corresponding couplings
\begin{align}
    \label{eq:SUNpot}
    V(\phi_\text{adj}) = \lambda_{1}\left(\text{Tr}\,\phi_\text{adj}^2\right)^2 + \lambda_{2}\,\text{Tr}\,\phi_\text{adj}^4 
    \;.
\end{align}
The corresponding one-loop $\beta$-functions can be found in \cite{Einhorn:2017jbs}, and, upon addition of the gravitational contribution $\sim A$, are given by
\begin{align}
\label{eq:SUNbetas-1}
    \beta_{\lambda_1} &= \frac{N^2+7}{16 \pi ^2}\lambda _1^2+\frac{3(N^2+3)}{4 \pi ^2 N^2}\lambda _2^2 +\frac{(2N^2-3)}{4 \pi ^2 N}\lambda _1 \lambda _2    
    -\frac{3N \alpha}{\pi }\lambda_1+18 \alpha ^2+A\,\lambda _1 
    \;,
    \\
    \label{eq:SUNbetas-2}
    \beta_{\lambda_2} &= +\frac{N^2-9}{4 \pi ^2 N}\lambda _2^2+\frac{3}{4 \pi ^2}\lambda _1 \lambda _2-\frac{3N \alpha}{\pi }\lambda _2
    +3N \alpha ^2
    +A\,\lambda _2
    \;.
\end{align}
Whenever real fixed points exist, these include a fully IR-attractive fixed point which obeys
\begin{align}
    \lambda_{1,\,\ast} < \lambda_{2,\,\ast} <0\;.\label{ineq:fps}
\end{align}
For the phenomenologically interesting case of $N=5$, real fixed points exist for $A\gtrsim 10 \alpha_{\ast}$. For instance, the case $\alpha_\ast=1/40$ and $A=2$ lies well within this region. This case is of interest, as $\alpha \approx 1/40$ is the approximate value of the Standard-Model gauge couplings in the region where they nearly cross, and $A=2$ is, as we will see, a critical value when it comes to the relevance of the mass parameter.\\
In Fig.~\ref{fig:adjointSU5_running} we show an explicit example for the ``focusing" of arbitrary initial conditions to the  IR-attractive fixed point for $\lambda_{1/2}$ that occurs as a consequence of the irrelevance of the couplings.  In other words,  large UV ranges of couplings are mapped to much smaller IR ranges. For a true fixed-point trajectory, the IR ranges become points, i.e., $\lambda_{1,2}(k\gtrsim M_{P})=\lambda_{1,2\, \ast}$.
This exemplifies the increase in predictivity for all quartic couplings.\\
The quartic couplings do, however, not remain at their fixed-point values for all scales, but start to deviate for $k< M_P$. This is due to the running of $A$ that we will explain in detail below. In short, $A=\rm const$ holds in the fixed-point regime.  At lower scales, it decreases very quickly and can then be neglected for the running of the quartic couplings.
The remaining contributions from matter fluctuations in Eqs.~\eqref{eq:SUNbetas-1}-\eqref{eq:SUNbetas-2} drive the quartic scalar couplings away from the fixed-point value. In summary, asymptotic safety  predicts the ``initial values'' for the flow at $k$ below $M_P$.\\

In the graviton approximation (neglecting scalar fluctuations of the metric), and using a Litim-type cutoff \cite{Litim:2001up}, the anomalous dimension $A$ is given by \cite{Wetterich:2017ixo}, see also \cite{Narain:2009fy,Percacci:2015wwa,Eichhorn:2017als},
\begin{align}
    \label{eq:quartic_spin-2}
    A = \frac{5}{12\pi^2\,\tilde{M}_\text{P}^2(1-v_0)^2}\;.
\end{align}
Here, $\tilde{M}_\text{P} = M_\text{P}(k)/k$ is the dimensionless flowing Planck mass while $v_0=2U_0/(M_\text{P}^2k^2)$ involves the value of the effective scalar potential $U$ at the minimum, related to the cosmological constant. The full gravitational contribution receives an additional correction from the scalar fluctuation of the metric \cite{Pawlowski:2018ixd}. It contributes less than $5\%$ of the graviton (traceless tensor) fluctuations. Within a truncation of a flowing field-dependent potential and Planck mass, $A$ is found to be of the order one for GUTs \cite{Wetterich:2019zdo}. Significant systematic uncertainties remain in the quantitative determination of the fixed-point value of $A$, which depends on the gravitational fixed-point values under the impact of matter, see, e.g., \cite{Dona:2013qba, Meibohm:2015twa,Biemans:2016rvp,Eichhorn:2018nda}. For the predictivity for quartic scalar couplings, only the sign of $A$ plays a role, whereas predictivity for mass parameters additionally depends on the value of $A$. We stress that $A$ is universal for all quartic couplings of scalar fields as gravity is ``blind" to internal symmetries. This property was first exploited in \cite{Eichhorn:2017als} to derive that the Higgs portal to uncharged scalar dark matter is dynamically driven to zero under the impact of quantum gravity.
\\

The flow of $A$ is essentially determined by the flow of the dimensionless squared Planck mass $\tilde{M}_\text{P}^2$ which is a relevant parameter. 
Therefore, the crossover scale $k_t$ from the fixed-point regime to the regime where $\tilde{M}_P^2$ runs and quantum scale invariance does not hold, can be chosen freely. Different choices of $k_t$ result in different low-energy values of the dimensionful running Planck mass $M_P(k)$ at $k\ll k_t$. Indeed, the flowing dimensionful squared Planck mass
\begin{align}
    \label{eq:running-Planck-mass-2}
    M_P^2(k) = \tilde{M}_P^2(k)\,k^2 = M_P^2\left(1+\frac{k^2}{k_t^2}\right) 
\end{align}
freezes for $k\ll k_t$, which is the range of classical gravity.  Here, 
the dimensionful squared Planck mass $M_P^2$ (distinguished from its \emph{running} counterpart $M_P(k)$ by the lack of argument) 
corresponds to the presently observed value. 
 To recover this low-energy value of the dimensionful Planck mass in the IR, $M_P(k=0)=M_P$, one typically has to choose $k_t \approx M_P$, as we have already implicitly assumed in the discussion above. The precise 
relation between $M_P$ and the transition scale $k_t$,
\begin{align}
    k_t = \frac{M_P}{\tilde{M}_{P\,\ast}}\;,\label{eq:ktdef}
\end{align}
depends on the fixed-point value of $\tilde{M}_{P\,\ast}$ at the Reuter fixed point. Actually, $k_t$ may be substantially smaller or larger than $M_P$ for fixed-point values significantly different from 1.\\

Below the crossover scale $k_t$, the dimensionless coupling $\tilde{M}_{P\ast}$ starts running fast, with the approximate solution to the flow equation given by
\begin{align}
    \label{eq:running-Planck-mass}
    \tilde{M}_P^2(k) = \tilde{M}_{P\ast}^2\left(\frac{k_t^2}{k^2} + 1\right) = \frac{M_P^2}{k^2} + \tilde{M}_{P\ast}^2\;,
\end{align}
as already implied by Eq.~\eqref{eq:running-Planck-mass-2} and Eq.~\eqref{eq:ktdef}.
The inverse dimensionless coupling $1/\tilde{M}_P^2(k)$ can be taken as a measure of the strength of gravity. Below $k_t$, it decreases rapidly,  effectively decoupling gravitational fluctuations at lower momentum scales and leading to a very rapidly decreasing $A$ in this region.
As a result, the gravity-induced anomalous dimension $A$ is effectively only present above the transition scale $k_t$, where the dimensionless Planck mass $\tilde{M}_{P\ast}$ is constant due to the presence of the Reuter fixed point. \\

The 
constant value of $A$ for $k\gg k_t$
results in $\lambda_H(k)$ being given by the partial fixed-point value \eqref{eq:lambda-FP} for the whole range $k>k_t$. For $k<k_t$, the running of $\lambda_H$ follows the (perturbative) running of the low-energy effective theory, i.e., a GUT or the SM. With initial value 
$\lambda_H(k_t)$ given by \eqref{eq:lambda-FP},
the low-energy value becomes predictable, implying a prediction of the ratio of the mass of the Higgs boson to the Higgs vev in the SM and fixing the quartic couplings in GUTs at the GUT-breaking scale.
For practical computations\footnote{In practice, to determine the critical value of $\lambda_i(k)$ for $k\ll k_t$ that corresponds to the unique fixed-point trajectory, one can integrate the flow of $\lambda_i(k)$ from the IR to the UV; using nested intervals of IR values to find the unique IR value for which the trajectory reaches the fixed point for $k>k_t$.}, one may fix $\lambda_i(k_\text{in}) = \lambda_{i\;\ast}$ for $k_\text{in}$ sufficiently above $k_t$, e.g., $k_\text{in}=100\,k_t$.\\

For a detailed numerical solution of the flow equation \eqref{eq:chr-A}, or the corresponding expressions for GUT-couplings $\lambda_i$, cf.~Eqs.~\eqref{eq:SUNbetas-1}-\eqref{eq:SUNbetas-2}, we need, beyond the ``perturbative quantities'' $\mathcal{C}_g$, $\mathcal{C}_y$, $c_\lambda$ and $\eta$, the value of $A$ for every scale $k$. Within the uncertainties remaining in the quantum-gravity computations, one can use the approximate form \eqref{eq:quartic_spin-2}, with $\tilde{M}_P^2$ given by \eqref{eq:running-Planck-mass}, where the parameter $\tilde{M}_{P\,\ast}^2$ can be computed for a given GUT model, see, e.g., \cite{Dona:2013qba,Biemans:2017zca,Wetterich:2019zdo}. Let us stress that the existence of a gravitational fixed point is not guaranteed for an arbitrary matter content. Extended studies to determine the existence and location of the Reuter fixed point for a given GUT model with high precision are beyond the scope of this work.
For the running of $v_0(k)$ one may use
\begin{align}
    v_0(k) = \frac{2u_\ast}{\tilde{M}_P^2(k)}\;,
\end{align}
with constant $u_\ast$ again computable for a given GUT model \cite{Dona:2013qba,Biemans:2017zca,Pawlowski:2018ixd, Wetterich:2019zdo}. The detailed values of $\tilde{M}_{P\,\ast}^2$ and $u_\ast$ only play a minor role for the values of the quartic couplings at the GUT scale. The transition at the crossover scale $k_t$ is rather rapid as compared to the rather slow running of the couplings $g$, $y$, or $\lambda_i$. Predictions of $\lambda_i(k_t)$ only need the values of $\alpha(k_t)$ and $\alpha_y(k_t)$ which, in turn, can be extrapolated perturbatively for a given GUT-model from the measured low-energy values of the gauge couplings and the top-quark mass. \\

The predictions for $\lambda_i(k_t)$ do not depend much on the fixed-point values of $g$ and $y$. Either $g(k_t)$ and $y(k_t)$ could be given by non-vanishing values $g_\ast$ and $y_\ast$ of the UV fixed point \cite{Eichhorn:2017muy}, as first explored for SM couplings \cite{Harst:2011zx,Eichhorn:2017ylw,Eichhorn:2017lry,Eichhorn:2018whv}. 
Alternatively, the fixed-point values may vanish, $g_\ast = y_\ast = 0$, and the slow evolution of the relevant couplings $g$ and $y$ in the region $k>k_t$ could induce non-zero $g(k_t)$ and $y(k_t)$. 
A gravitationally induced interacting fixed point for $g$ and/or $y$ is always induced together with an asymptotically free fixed point at which the corresponding coupling is a relevant parameter.
For both cases the values $|\lambda_i(k_t)|$ turn out to be rather close to zero if $A$ is substantially larger than zero. This is sufficient for a prediction of $\lambda_i$ at the GUT-scale $k_\text{GUT}$, in particular if $k_\text{GUT}$ is substantially smaller that $k_t$. In this case, $\lambda_i(k_\text{GUT})$ will be determined by the perturbative flow within the GUT without gravity in the range $k_\text{GUT}<k<k_t$. 
 \\

The sign of the quartic couplings $\lambda_i(k_t)$ depends on the relative size of the gauge- and Yukawa contributions. One typically finds a negative coupling $\lambda_i(k_t)<0$
whenever gauge fluctuations dominate, $\lambda_{i,\,\ast}<0$. 
In contrast, if the Yukawa fluctuations dominate,
one obtains $\lambda_i(k_t)>0$. We emphasize that a negative value of $\lambda_i(k_t)$ is not necessarily a sign of instability. The quartic couplings only capture an expansion of the potential around vanishing values of the scalar field. The behavior for large values of the scalar fields may be very different since the scalar potential near the fixed point is typically not of a polynomial form \cite{Wetterich:2019qzx}. 
In our discussion, we have neglected non-minimal scalar-gravity couplings $\xi$. They typically add to the RHS of Eq.~\ref{eq:chr-A} a term \cite{Wetterich:2019qzx}
\begin{align}
    \mathcal{C}_\xi = -\frac{8\xi^2\,v}{\tilde{M}_P^2}\;.
\end{align}
For small $\xi$, that we assume here, this term induces only minor modifications for the prediction of quartic couplings at the GUT-scale. For larger $\xi$ these predictions require knowledge of $\mathcal{C}_\xi$. The possibility to predict the quartic couplings is not affected if $\lambda$ remains an irrelevant coupling, but the prediction may depend on other parameters. Similarly, could have an important impact on Eq.~\ref{eq:chr-A}, but is typically itself an irrelevant, and thus predictable, quantity. 
\\

\section{Flow of the effective potential for a gauge-Yukawa model}
\label{sec:simple-U(1)}
The basic features of the quantum-gravity predictions for effective scalar potentials can be understood in the simplest model with gauge and Yukawa interactions. Here, we discuss 
an abelian $U(1)$-gauge theory with a charged scalar described by a single complex field $\phi$.
We add a neutral Dirac fermion where left- and right-handed components $\psi_{L,\,R}$ couple
to the complex scalar via a Yukawa term 
$y\,(\bar{\psi}_R\,\phi^*
\,\psi_L - \bar{\psi}_L\,\phi
\,\psi_R)$
with Yukawa coupling 
$y$. While this simple 
model lacks a non-trivial group structure, it exemplifies how an asymptotically safe fixed point determines the scalar quartic couplings, higher-order scalar self-interactions, and potentially even the corresponding symmetry-breaking scale $M_X$.\\

We employ the gauge-invariant flow equation \cite{Wetterich:2016ewc} or, equivalently, a background-field formalism in the linear split with so-called physical gauge fixing. In this setting, the gauge modes decouple from the gauge-invariant modes and only generate a field independent term $\tilde{\eta}$.
The flow of the dimensionless scalar potential $u = U/k^4$ at a fixed dimensionless field variable 
$\rho = \phi^*\phi / k^2$ is given by
\begin{align}
    k\,\partial_k\,u(\rho) &=
    -4 u(\rho) + (2+\eta_\phi)\rho\,u'(\rho) 
    + \tilde{\pi}_g
    + \tilde{\pi}_f
    + \tilde{\pi}_2 + \tilde{\pi}_0 + \tilde{\eta} \;,
    \label{eq:U1-complex-scalar-potential}
\end{align}
with primes denoting derivatives with respect to $\rho$. The contributions from gauge bosons $\tilde{\pi}_g$ and fermions $\tilde{\pi}_f$ are the standard contributions in flat space without gravity, as computed in early investigations of functional renormalization for scalar potentials \cite{Wetterich:1990an, Reuter:1993kw}. They have been investigated in the context of chiral Yukawa systems 
, e.g., in 
\cite{
Gies:2009sv}. Note that these calculations also contain contributions of gauge and fermion fluctuations to the scalar anomalous dimension $\eta_\phi$. To obtain the universal one-loop beta functions for the GUT without gravity, the standard perturbative one-loop expressions for the anomalous dimension have to be taken into account. These can 
be 
obtained from the flow of the scalar wave-function renormalization (coefficient of the scalar kinetic term) in the FRG framework. We do not display the derivation of $\eta_\phi$ here and simply take the one-loop perturbative values for this  quantitatively very small effect, thereby neglecting threshold-corrections that occur, e.g., at the onset of spontaneous symmetry breaking.\\

The graviton contribution (TT-contribution) $\tilde{\pi}_2$ \cite{Narain:2009fy, Wetterich:2017ixo, Eichhorn:2017als} is given for the Litim cutoff function \cite{Litim:2001up}
as
\begin{align}
    \label{eq:spin-2-contributions}
    \tilde{\pi}_2 &= \frac{5}{24\pi^2}\frac{1}{1-v(\rho)}\;,\quad v(\rho)=2u(\rho)/\tilde{M}_P^2\;.
\end{align}

The scalar contribution $\tilde{\pi}_0$ \cite{Pawlowski:2018ixd} mixes the fluctuations of the complex scalar field $\phi$ and the gravitational spin-0 mode.
It also includes self-interactions of the physical scalar
\begin{align}
    \label{eq:spin-0-contributions}
        \tilde{\pi}_0 &= \frac{1}{24\pi^2}\frac{\left(1+u'(\rho) + 2\rho\,u''(\rho)\right) + \frac{3}{4}\left(1-v(\rho)/4\right)}{\left(1-v(\rho)/4\right)\left(1+u'(\rho) + 2\rho\,u''(\rho)\right) + 3\rho\,u'(\rho)^2/\tilde{M}_P^2}\;.
\end{align}
The measure contribution $\tilde{\eta} = -1/(8\pi^2)$ also includes the effect of fluctuations of the gravitational gauge degrees of freedom and the flow of the renormalized Faddeev-Popov determinant or ghost terms. It does not depend on $\rho$ and only contributes to the flow of the cosmological constant, providing for the correct counting of the physical degrees of freedom.\\

Finally, the gauge boson and fermion contributions read,
\begin{align}
    \label{eq:gauge-and-Yukawa-contributions}
    \tilde{\pi}_g = \frac{1}{32\pi^2}\left(\frac{3}{1+g^2\rho}
    - 1\right)
    \;,\quad\quad\quad
    \tilde{\pi}_f = -\frac{1}{8\pi^2}\frac{1}{1+y^2\rho}    
    \;,
\end{align}
where the constant term in $\tilde{\pi}_g$ arises as the measure contribution in the U(1)-gauge sector.
Eq.~\eqref{eq:gauge-and-Yukawa-contributions} highlights that gauge and Yukawa couplings contribute with opposite signs. 
We will come back to a numerical solution of Eq.~\eqref{eq:U1-complex-scalar-potential} for the full potential when analyzing the implications of an asymptotically safe fixed point for spontaneous symmetry breaking in Sec.~\ref{sec:SSB-U1}.\\

The mixing between $\phi$ and the metric scalar fluctuation is a subdominant effect that may be neglected for an analytic discussion. If we omit the last term in the denominator in Eq.~\eqref{eq:spin-0-contributions}, the scalar contribution consists of separate contributions $\tilde{\pi}_{0,\,\phi}$ from $\phi$-fluctuations and $\tilde{\pi}_{0,\,g}$ from metric fluctuations,
\begin{align}
    \tilde{\pi}_{0,\,\phi} &= \frac{1}{32\pi^2}\left(\frac{1}{1+u'(\rho)+2\rho u''(\rho)} + \frac{1}{1+u'(\rho)}\right)\;,
    \\
    \tilde{\pi}_{0,\,g} &= \frac{1}{24\pi^2(1-v(\rho)/4)}\;.
\end{align}
For the gauge-invariant flow equation, the flow of the $\rho$-derivative of $u$ is obtained directly as the $\rho$-derivative of the rhs in Eq.~\eqref{eq:U1-complex-scalar-potential}. Neglecting $\eta_\phi$, one finds 
\begin{align}
    k\,\partial_k\,u'(\rho)\; =& \;\left(A(\rho) - 2\right)u'(\rho) + 2\rho\,u''(\rho)
    - \frac{A(\rho)v(\rho)\xi(\rho)}{2} - \frac{3g^2}{32\pi^2(1+g^2\rho)^2} + \frac{y^2}{8\pi^2(1+y^2\rho)^2}
    \notag\\
    &- \frac{1}{32\pi^2}\left(\frac{3u''(\rho)+2\rho\,u'''(\rho)}{(1+u'(\rho)+2\rho\,u''(\rho))^2} + \frac{u''(\rho)}{(1+u'(\rho))^2}\right)\;,
    \label{eq:scale-der-of-U-in-gauge-Yukawa-model}
\end{align}
with
\begin{align}
    A(\rho) = \frac{5}{12\pi^2\,\tilde{M}_P^2(1-v(\rho))^2} + \frac{1}{48\pi^2\,\tilde{M}_P^2(1-v(\rho)/4)^2}
    \quad\text{and}\quad
    \xi(\rho) = \frac{\partial\,\tilde{M}_P^2}{\partial\rho}\;.
\end{align}
Here, we have accounted for a $\rho$-dependence of $\tilde{M}_P$, promoting the coupling to a field-dependent function, which is simply a way of including non-minimal interactions, e.g., $\xi(\rho)\,\phi^{\ast}\phi\, R$. In the following, we simplify by omitting the $\rho$-dependence of $\tilde{M}_P^2$, setting $\xi=0$.\\

To obtain the quartic coupling, we need a further $\rho$-derivative of Eq.~\ref{eq:scale-der-of-U-in-gauge-Yukawa-model}.  We further omit the second term in the gravity-induced anomalous dimension $A$ which is much smaller than the first term. If we also neglect higher-order terms $u'''$ and $u^{(4)}$, one finds
\begin{align}
    \label{eq:scalar-quartic-from-potential-U1}
    k\,\partial_k\,u'' = A\,u'' + (\partial_\rho A)\,u' + \frac{3g^4}{16\pi^2(1+g^2\,\rho)^3} - \frac{y^4}{4\pi^2(1+y^2\,\rho)^3} + \frac{1}{16\pi^2}\left(\frac{9\,u''^2}{(1+u'+2\rho\,u'')^3} + \frac{u''^2}{(1+u')^3}\right)\;.
\end{align}
We define $\lambda = u''(0)$ and neglect a possible small mass term $\tilde{m}_\phi^2 = u'(0)$. Evaluated at $\rho=0$, Eq.~\eqref{eq:scalar-quartic-from-potential-U1} reduces to Eq.~\eqref{eq:chr-A}, with
\begin{align}
    \mathcal{C}_g = \frac{3g^4}{16\pi^2}\;,\quad\quad
    \mathcal{C}_y = \frac{y^4}{4\pi^2}\;,\quad\quad
    c_\lambda = \frac{5}{8\pi^2}\;.
\end{align}

\section{Spontaneous symmetry breaking}
\label{sec:SSB}
Spontaneous symmetry breaking (SSB) occurs once the origin of field space becomes a local maximum, signaled by negative quadratic terms in the expansion of the potential around the origin.
The scale of spontaneous breaking, therefore, follows from the behavior of the scalar mass-terms $m_j^2$. 
More precisely, expectation values of the scalar fields are typically determined as $\sqrt{m_j^2/\tilde{\lambda}_j}$, with $\tilde{\lambda}_j$ an ``effective quartic scalar coupling'' which is of the characteristic size of the quartic couplings.
Within an effective-field theory treatment, the values of $m_j^2$ are free parameters. Thus, it is a matter of choice, whether SSB occurs at all, and what the corresponding scale is.
Within an asymptotically safe UV completion of a GUT, the mass-terms can become irrelevant if gravity fluctuations have sufficient strength. In that case, the fixed-point values of the masses and the ensuing flow determine uniquely whether SSB occurs or not. Moreover, the scale of SSB is no longer a matter of choice. \\

Whether the dimensionless couplings $\tilde{m}_j^2 = m_j^2/k^2$ are relevant or irrelevant parameters at the UV fixed point, essentially depends on the value of $A$. Indeed, the same universal gravity-induced anomalous dimension appears both in the flow equation of the quartic couplings and the mass terms. This can be easily seen from Eq.~\eqref{eq:scale-der-of-U-in-gauge-Yukawa-model} if we define $\tilde{m}^2 = u'(\rho=0)$. 
The flow of a generic scalar mass-term $\tilde{m}_\phi^2$ is governed by its $\beta$-function
\begin{align}
\label{eq:mass-beta-schematic}
    \beta_{\tilde{m}_\phi^2} = k\,\partial_k\,\tilde{m}_\phi^2 = \left(A-2+\eta \right)\tilde{m}_{\phi}^2
    +\mathcal{C}_y^m(\alpha_y)
    -\mathcal{C}_g^m(\alpha)
    +\dots
\end{align}
It includes a canonical term $-2\,\tilde{m}_\phi^2$, the anomalous dimension due to quantum effects of the gravitational field $A$, and explicit contributions due to gauge and fermion fluctuations, as well as a quantitatively negligible non-gravitational anomalous dimension $\eta$.
The contributions $\mathcal{C}_g^m$ and $\mathcal{C}_y^m$ can be inferred from equations of the type of Eq.~\eqref{eq:scale-der-of-U-in-gauge-Yukawa-model} for $\rho=0$. The vanishing of the rhs of Eq.~\eqref{eq:mass-beta-schematic} defines the ``critical surface'' \cite{Wetterich:1983bi,Aoki:2012xs}.
For clarity and as it does not change our main line of argument, we  
will again neglect
gauge-Yukawa contributions 
$\eta$
to the anomalous dimension and have omitted additional contributions due to scalar self-interactions. They will be included in numerical calculations later on.\\

The running of the scalar mass vanishes for the fixed point or scaling solution
\begin{align}
\label{eq:mFP}
    \tilde{m}^2_{\phi,\,\ast} = \frac{\mathcal{C}_y^m(\alpha_y)-\mathcal{C}_g^m(\alpha)}{2-
    A
    }\;.
\end{align}
On the "critical surface" given by the scaling relation Eq.~\eqref{eq:mFP}, the dimensionful mass term $m_{\phi}^2(k) = \tilde{m}_{\phi}^2\,k^2$ vanishes for $k=0$. Neglecting mixing with other couplings, the critical exponent for small deviations from the scaling solution Eq.~\eqref{eq:mFP} is given by
\begin{align}
    \theta_{\tilde{m}^2_\phi} = -\partial_{\tilde{m}^2_\phi}\beta_{\tilde{m}_\phi^2}\Big|_{\tilde{m}^2_\phi=\tilde{m}^2_{\phi,\,\ast}} = 2-
    A
    \;.
\end{align}
Gravitational fluctuations render the mass terms less relevant 
since the canonical dimension two is replaced by $2-A$. This even holds in the case where the fixed-point value is vanishing, i.e., $\tilde{m}^2_{\phi,\,\ast}=0$.
Whenever the gravitational contribution dominates over the canonical term, i.e., 
$A>2$, the fixed point is IR-attractive. Then, the model realizes  self-organized criticality \cite{Bornholdt:1992up, Gies:2013pma} and the mass-term becomes a prediction \cite{Wetterich:1981ir,Wetterich:2016uxm}. In this case, $\tilde{m}_j^2(k_t)$ equals the fixed-point value $\tilde{m}_{j\,\ast}^2$. 
In the opposite case, i.e., if 
$A<2$, the mass-term is relevant and hence a free parameter of the theory. The low-energy values of relevant parameters cannot be predicted, even though their fixed-point values are of course calculable. We can, therefore, set the values $\tilde{m}_j^2(k_t)$ as free parameters in this case. In this scenario, the RG-trajectory of the scalar mass-term 
can flow to positive or negative values, independently of the fixed-point value of $\tilde{m}_{\phi\,\ast}^2$. \\

This divides the gravitational parameter space into a ``strong-gravity" and a ``weak-gravity" regime, as $A>2$ can only be reached if the effective coupling of metric fluctuations (which depends on the running  Planck mass but also additional gravitational couplings) is large enough. The determination of the corresponding boundary in the Einstein-Hilbert truncation has been undertaken in \cite{Eichhorn:2017als,Pawlowski:2018ixd}.\\

The fixed-point value in Eq.~\eqref{eq:mFP} is positive (negative) whenever gauge (Yukawa) contributions are dominant. 
This hints at the intriguing possibility that asymptotic safety might automatically result in $M_{\rm GUT}\lesssim M_P$ for a suitable choice of scalar multiplets.
\begin{table}[!t]
\centering
\begin{tabular}{c|c||c|c}
    $A$ & $\mathcal{C}_y\lessgtr\mathcal{C}_g$ & $m_{\phi,\,\ast}^2$ & $\lambda_{4,\,\ast}$
    \\
    \hline
    \hline
    $A> 2$ & $\mathcal{C}_y>\mathcal{C}_g$ & $m_{\phi,\,\ast}^2<0$ & $\lambda_{4,\,\ast}>0$
    \\
    \hline
    $A > 2$ & $\mathcal{C}_y<\mathcal{C}_g$ & $m_{\phi,\,\ast}^2>0$ & $\lambda_{4,\,\ast}<0$
    \\
    \hline\hline
    $A< 2$ & $\mathcal{C}_y>\mathcal{C}_g$ & $m_{\phi,\,\ast}^2>0$ & $\lambda_{4,\,\ast}>0$
    \\
    \hline
    $A < 2$ & $\mathcal{C}_y<\mathcal{C}_g$ & $m_{\phi,\,\ast}^2<0$ & $\lambda_{4,\,\ast}<0$
\end{tabular}
\caption{\label{tab:stability-cases}
    The qualitatively different fixed-point structure for the mass-term and quartic coupling $\lambda_4$ of the scalar potential depends on gravitational ($A > 2$), gauge ($\mathcal{C}_g$) and Yukawa ($\mathcal{C}_y$) contributions at the UV fixed point.
}
\end{table}
The generic fixed-point structure for quartic couplings $\lambda_i$ and masses $m_i$ is summarized in Tab.~\ref{tab:stability-cases}. We repeat that no conclusions should be drawn about the global stability of the potential in cases where an instability around the origin is present, as this can simply be an indication of spontaneous symmetry breaking (with a globally stable potential) in the fixed-point regime.\\

We first consider the case
$0<A<2$, where the mass term remains a free parameter. For $\mathcal{C}_y^m(\alpha_y)>\mathcal{C}_g^m(\alpha)$, one infers a stable scaling potential near the origin with a local minimum at vanishing field value. Within the gauge-Yukawa model of Sec.~\ref{sec:simple-U(1)}, we will explore the fixed-point potential to higher orders in a polynomial expansion in Sec.~\ref{sec:SSB-U1_stab}, to confirm that the potential is indeed stable near the origin in field space.
SSB can occur due to the flow of $\tilde{m}_\phi^2$ away from the fixed-point value. The expectation value depends on the 
free parameter corresponding to the mass term, $\tilde{m}_{\phi}(k_t)$. 
Below the transition scale $k_t$, gravitational contributions switch off dynamically. Towards the IR, the scalar mass-term is driven to smaller values by the overall positive contribution $\mathcal{C}_y^m(\alpha_y)-\mathcal{C}_g^m(\alpha)$ in its $\beta$-function. Eventually, the mass crosses zero and dynamical SSB occurs. The scale at which this happens is determined by the free parameter $\tilde{m}_{\phi}(k_t)$. There is a critical value of $\tilde{m}_{\phi}(k_t)$ for which SSB occurs, dividing the parameter space into a symmetric and a symmetry-broken phase.
The remaining shape of the scalar potential is predicted completely by the asymptotically safe fixed point without involving any other free parameters. Therefore, even though the scale of SSB is a choice, the direction of SSB in settings with physically different vacua is determined.
\\

The fully predictive case, i.e., $A>2$, is phenomenologically even more appealing because it would allow to not only determine the direction of the scalar vacuum expectation value, but also the symmetry-breaking scale. While we are able to construct such trajectories within simple truncations of the scalar potential, questions of global stability of the scalar potential remain inconclusive. Hence, we defer this case to later studies. Nevertheless, we anticipate, that if such a setting admits a fully stable potential, the associated GUT symmetry-breaking scale will be predicted to lie in the vicinity of the Planck scale. This is a simple consequence of the fact that fixed-point values are roughly of the order or below one. This fixes the dimensionless vev in the fixed-point regime to be roughly of order one, translating into a dimensionful vacuum expectation value slightly below the Planck scale. Thus, this case could provide a dynamical explanation for the non-observation of proton decay.
\\

To find all the possible cases for the scalar potential, we have, until now, treated the fixed-point values $\alpha_{g,\,\ast}$ and $\alpha_{y,\,\ast}$ as free parameters. In a specific GUT, these couplings are typically also fixed (or at least bounded from above) by the transplanckian UV-scaling regime.
In particular, as we have recently proposed in \cite{Eichhorn:2017muy}, based on studies of the gravitational effects in \cite{Harst:2011zx, Eichhorn:2017lry}, the unified gauge coupling could indeed be predicted as a consequence of asymptotically safe fixed-point scaling. The physical mechanism behind this proposal is based on the interplay of quantum-gravity fluctuations and matter fluctuations. This forces the gauge coupling towards one unique trajectory that is singled out by demanding quantum scale-invariance in the UV, as the gauge coupling becomes an irrelevant parameter in this setting.
In more detail, the $k$-dependence of the gauge coupling in the presence of quantum-gravity fluctuations is given by
\begin{align}
k\partial_k\, \alpha_g=\beta_{\alpha_g}=\eta_{g} \alpha_g+\left(\mathcal{N}-
\mathcal{N}_c\right)\frac{\alpha_g^2}{4\pi}\;.
\end{align}
Here $\mathcal{N}-\mathcal{N}_c$ accounts for the fluctuation contribution of gauge bosons, fermions and scalars.
With enough additional matter content  
for this contribution to screen (instead of antiscreen) the gauge coupling, i.e., $\mathcal{N}>\mathcal{N}_c$, the gauge coupling $\alpha_g$ could become asymptotically safe under the influence of sufficiently strong gravitational fluctuations encoded in the negative value $\eta_g$ (the analogue of $A$), cf.~\cite{Eichhorn:2017muy,Harst:2011zx, Eichhorn:2017lry}. Indeed, the gravitational contribution $\eta_g(\tilde{M}_P)$ has been found to be negative, $\eta_{g}\leq 0$, 
in all studies up to date \cite{Daum:2009dn,Folkerts:2011jz,Harst:2011zx,Christiansen:2017gtg,Eichhorn:2017lry,Christiansen:2017cxa} in the Einstein-Hilbert truncation, see \cite{deBrito:2019umw} for the extension to higher-order gravitational couplings. 
The non-vanishing fixed-point value for the gauge coupling is determined solely by the group-theoretic data $\mathcal{N}-\mathcal{N}_c$ and the gravitational contribution $\eta_g$, i.e.,
\begin{align}
    \alpha_{g,\,\ast} = -\frac{4\pi\,\eta_g}{\mathcal{N}- \mathcal{N}_c}\;.
    \label{eq:alphafp}
\end{align}
At this fixed point, the gauge coupling $\alpha_{g,\,\ast}$ is IR attractive and thus a prediction of asymptotically safe fixed-point scaling, cf.~\cite{Eichhorn:2017muy}. In simple terms, quantum-gravity fluctuations drive the gauge coupling towards $\alpha_{g,\,\ast}$, resulting in the unique prediction 
$\alpha(k_t)= \alpha_{g,\,\ast}$, even if initial conditions for the RG flow in the far ultraviolet are chosen away from $\alpha_{g,\,\ast}$. Similar IR-attractive fixed points can be anticipated, cf.~\cite{Eichhorn:2017ylw,Eichhorn:2018whv}, to also be present for the Yukawa couplings of a specific GUT-model. Again, they will be determined solely in terms of the group-theoretic structure and the quantum-gravity contribution.
\\

Overall, the following tentative model-building scenario emerges: If gravity should indeed be asymptotically safe under the inclusion of all matter and gauge field fluctuations present in a given GUT model, the corresponding fixed-point regime could entirely fix all the gauge-, Yukawa- and quartic couplings of a given GUT model. By specifying the degrees of freedom and 
choosing the most predictive fixed point for gauge and Yukawa couplings, the 
IR-physics is determined in terms of a rather small set of free parameters. Thereby the predictive power of GUTs is dramatically enhanced and most GUTs can probably be excluded under the theoretical paradigm of the asymptotically safe UV fixed point.

\section{Spontaneous symmetry breaking for the gauge-Yukawa model}
\label{sec:SSB-U1}
\begin{figure}[!t]
\centering
\includegraphics[width=0.46\linewidth]{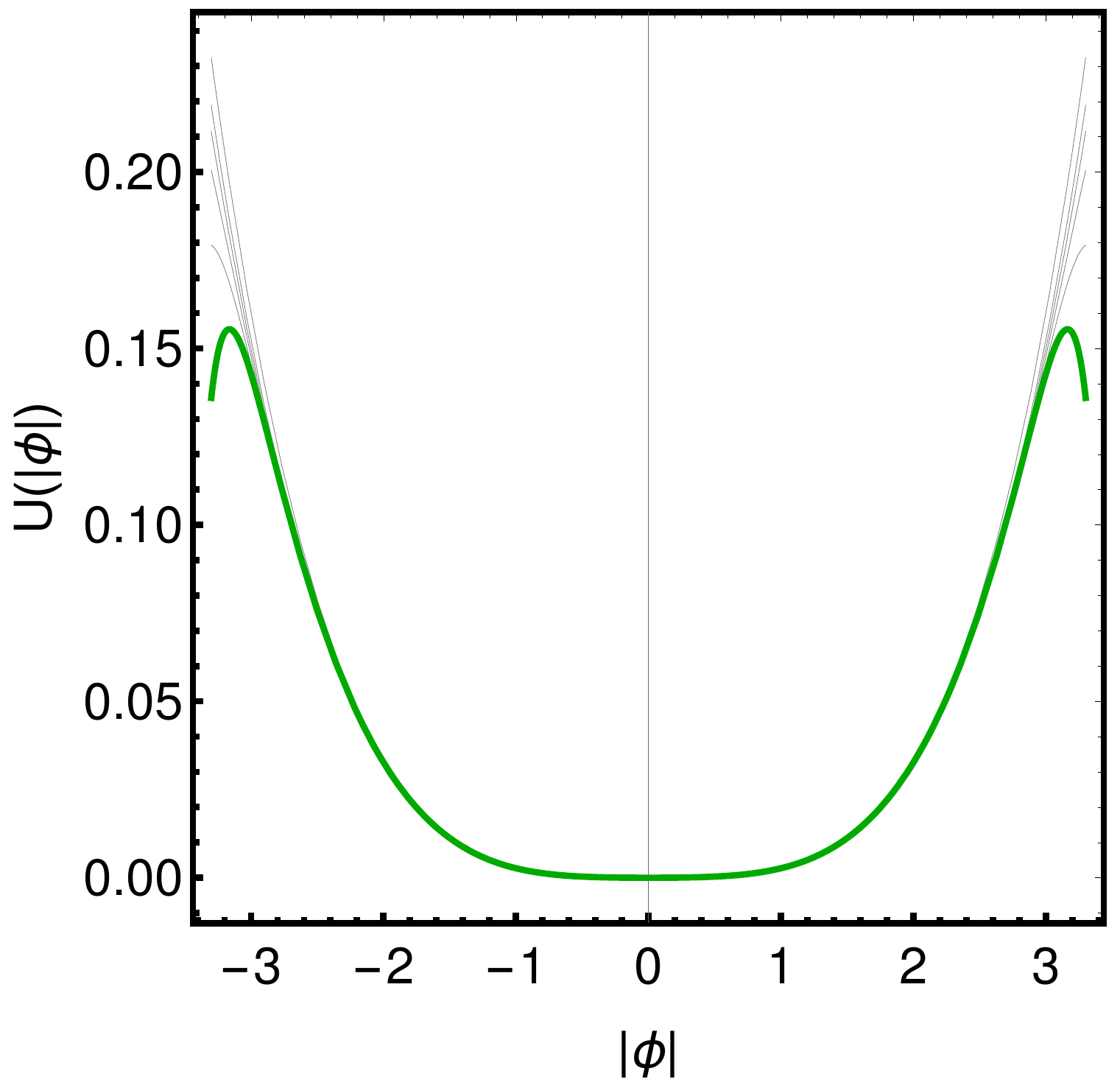}
\hfill
\includegraphics[width=0.467\linewidth]{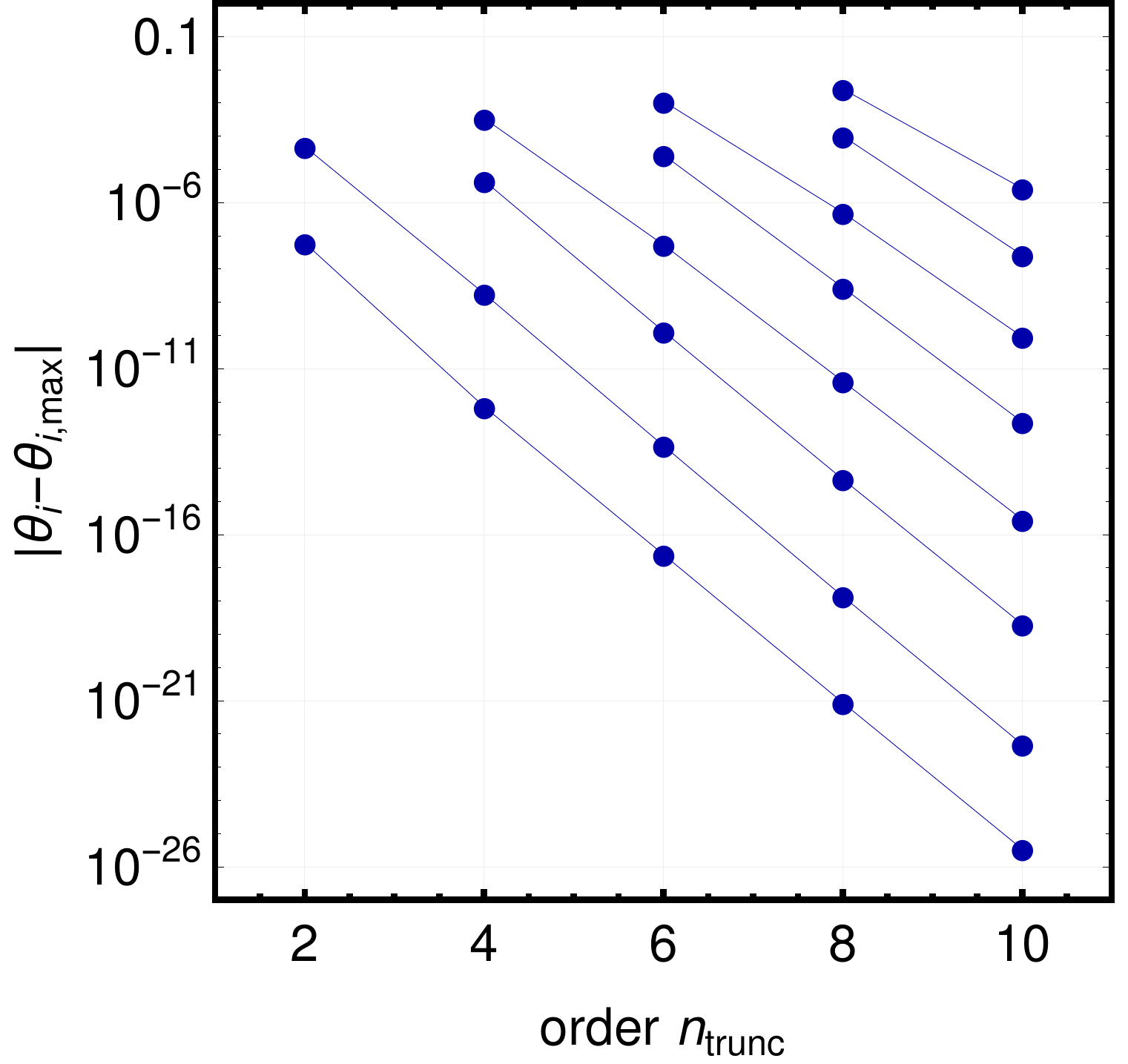}
\caption{\label{fig:U1-convergence}
Convergence of the scalar fixed-point potential for $A=1/10$, $\alpha_g=0$ and $\alpha_y=1/100$. 
Left-hand panel: Stable scalar potential within the radius of convergence at order $n_{\rm trunc}=12$ (thick green line) and at all lower orders (thin gray lines).
Right-hand panel: Apparent convergence of the 8 lowest order critical exponents $\theta_i$. For each $\theta_i$ we show the absolute value of the difference between the value at the present truncation order $n_\text{trunc}$ and at the highest truncation order $n_{\rm trunc,\,max}=12$, i.e., $|\theta_i - \theta_\text{i,\,max}|$.
}
\end{figure}

In the following, we explicitly discuss the potential, cf.~Eq.~\eqref{eq:U1-complex-scalar-potential} of the simple U(1) gauge-Yukawa model introduced in Sec.~\ref{sec:simple-U(1)} to demonstrate how spontaneous symmetry breaking is induced by the interplay of gauge, Yukawa and gravitational dynamics. As we will see, the entire potential (apart from the mass term) is fully fixed due to the predictive power of the trans-Planckian fixed point. Since symmetry breaking in this simple U(1) example does not distinguish different directions, we cannot demonstrate further predictive power with this example. The latter requires different potential vacuum expectation values and will be subsequently discussed in Sec.~\ref{sec:vev-direction-predicted}.

\subsection{Stability of the U(1) complex scalar fixed-point potential}\label{sec:SSB-U1_stab}
To investigate the stability of the scalar fixed-point potential $U(\rho)$ close to the origin, we expand the potential in Eq.~\eqref{eq:U1-complex-scalar-potential} in a polynomial expansion in the symmetric regime, i.e., for the dimensionful potential
\begin{align}
    \label{eq:SYM-expansion}
    U(\rho) = \sum_{n=1}^{n_{\rm trunc}}\frac{\lambda_{2n}^\text{(SYM)}}{n!}\rho^n\;.
\end{align}
To check convergence, we expand up to order $n_{\rm trunc,\,max}=12$.
Note that we can only identify the fixed point for even $n_{\rm trunc}$. All couplings $\lambda_{2n}$ for $n\geqslant2$, i.e., apart from the mass term, correspond to irrelevant directions and we observe fast apparent convergence of the critical exponents with growing order of the expansion, cf.~right-hand panel in Fig.~\ref{fig:U1-convergence} for the example of $A=1/10$, $\alpha_g=0$, and $\alpha_y=1/100$.

Within the radius of convergence of this expansion, the stability analysis around the origin confirms the intuition summarized in Tab.~\ref{tab:stability-cases}. In the case of a relevant mass term, the potential is found to be locally stable whenever the Yukawa contributions in Eq.~\eqref{eq:gauge-and-Yukawa-contributions} dominate. In contrast, the polynomial expansion indicates a local instability around $\rho=0$ whenever gauge contributions overcome Yukawa contributions. 
The effect of gauge- and Yukawa interactions is exchanged in the case $A>2$ for which the mass term becomes irrelevant. 

However, we stress that a polynomial expansion is never sufficient to determine global stability-properties, due to its finite radius of convergence. In particular, it is insufficient to conclusively determine the global stability and a possible non-trivial minimum of the scalar fixed-point potential for the case with $A>2$.
Here, we focus on the locally stable potential for $A<2$ and a dominant Yukawa contribution instead. In this case, the polynomial expansion is sufficient to reveal the onset of spontaneous symmetry breaking.

\subsection{Spontaneous symmetry breaking of the U(1) complex scalar}

To exemplify how spontaneous symmetry-breaking occurs below the Planck scale, we evolve the scalar couplings towards lower scales starting with initial conditions in the asymptotically safe fixed-point regime. We focus on the less predictive case,  $0<A<2$, and assume that Yukawa couplings dominate, cf.~row three in Tab.~\ref{tab:stability-cases}. The fixed-point potential is stable and the critical exponents up to order $n_{\rm trunc}=12$ converge very quickly, cf.~Fig.~\ref{fig:U1-convergence}. 

\begin{figure}[!t]
\centering
\includegraphics[width=0.8\linewidth]{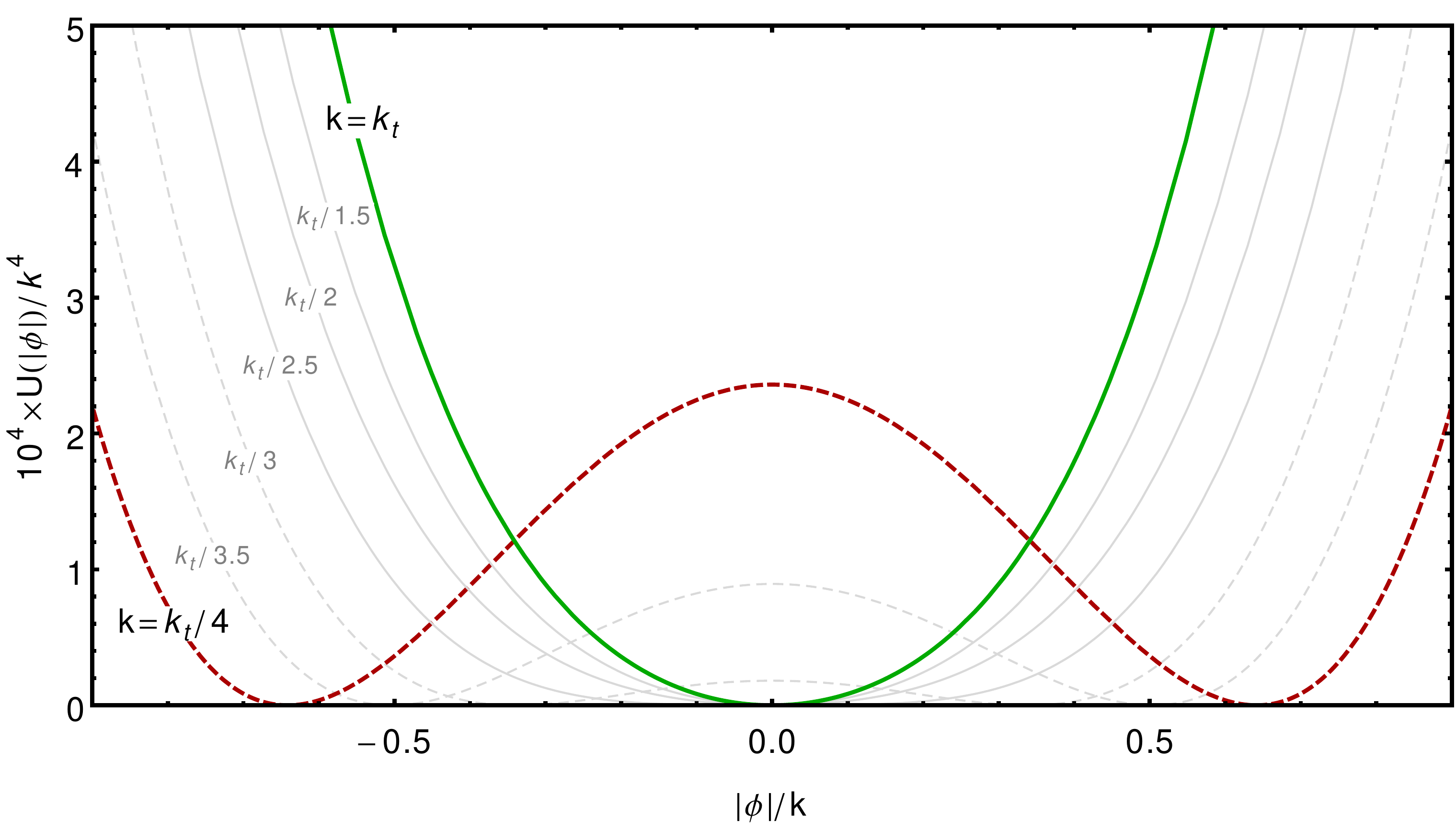}
\caption{\label{fig:U1-SSB-example}
Numerical evolution of the scalar potential, starting from the initial condition $U/k^4=u_{\ast}$ at $k=k_t$ (green continuous line).
Thin (dashed) lines depict the evolution before (and after) symmetry breaking. The latter occurs at $k/k_t\approx 1/2.5$. At $k/k_t\approx 1/4$ (red-dashed line) a non-trivial minimum has clearly developed. The gravitational, gauge and Yukawa contributions are chosen as $A=1/10$, $\alpha_g=0$ and $\alpha_y=1/100$. 
}
\end{figure}

Below the gravitational transition scale $k_t$, the dimensionless Planck-mass $\tilde{M}_{P}(k)$ grows 
quadratically and suppresses all gravitational contributions. 
The slow logarithmic running of gauge and Yukawa couplings is negligible. Thus we assume constant values for the latter.
The dominant Yukawa contributions drive the mass term to smaller values as long as they dominate over gauge and canonical contributions. As soon as the scalar mass term turns negative, we switch to a polynomial expansion around the non-trivial minimum $\kappa(k)$, i.e.,
\begin{align}
    U(\rho) = \sum_{n=2}^{n_{\rm trunc}}\frac{\lambda_{2n}^\text{(SSB)}}{n!}(\rho-\kappa)^n\;.
\end{align}
Fig.~\ref{fig:U1-SSB-example} shows how the fixed-point potential evolves to a symmetry-broken potential below the transition scale $k_t$.

Even though the mass term is relevant, we have chosen no deviation from its fixed-point value above $k_t$. Hence, no additional scale arises and symmetry breaking occurs very close to the transition scale. Note that the scalar mass term could be chosen to delay or even avoid any symmetry breaking. Since this arbitrariness is absent in the fully predictive case, it is of great interest to investigate the global stability of the fixed-point potential beyond polynomial expansions for these cases in the future.
\\

\section{Quantum-gravity predictions of spontaneous symmetry-breaking patterns}
\label{sec:vev-direction-predicted}

In more general GUT settings, the scalar  
multiplets
transform in a non-trivial representation of the GUT symmetry, which is typically reducible. Generically, the potential features more than one quartic coupling and symmetry breaking can a priori occur in different directions. Since the quartic couplings are fixed by gravitational fluctuations, the direction of symmetry breaking actually becomes a prediction of the asymptotically safe scaling regime. 
\\

In a general case with continuous or discrete symmetries excluding odd powers of the scalar fields $\mathbf{\Phi}_a$, a polynomial expansion of the quartic potential is given by
\begin{align}
    U(\mathbf{\Phi}) = \frac{1}{4!}\lambda_{abcd}\mathbf{\Phi}_a\mathbf{\Phi}_b\mathbf{\Phi}_c\mathbf{\Phi}_d + \text{mass terms} + \mathcal{O}(\mathbf{\Phi}^6)\;.
\end{align}
Neglecting contributions from non-marginal matter interactions and non-minimal couplings to gravity, the scale dependence of a generic scalar quartic coupling in this potential, denoted by $\lambda_{abcd}$, reads
\begin{align}
    \label{eq:quarticBeta}
  k\partial_k \lambda_{abcd}=   \beta_{\lambda_{abcd}} = \frac{1}{16\pi^2}\left(
    \mathbf{\Lambda}^2_{abcd}
    +\mathbf{S}_{2\; ae}^\text{(eff)}(\eta_\lambda,\alpha_g,\mathbf{Y})\,\lambda_{ebcd}
    + \mathbf{A}_{abcd}^\text{(eff)}(\alpha_g,\mathbf{Y}) \right)+ A\, \lambda_{abcd}\;,
\end{align}
see \cite{Machacek:1984zw}.
Here, $\mathbf{\Lambda}^2_{abcd} = \lambda_{abef}\lambda_{efcd} + \lambda_{acef}\lambda_{efbd} + \lambda_{adef}\lambda_{efbc}$, and we have again made use of the universality (independence from internal symmetries) of the gravitational contribution $\sim A$ , as given in Eq.~\eqref{eq:quartic_spin-2}.
The 1-loop contributions of gauge fields and fermions are universal (scheme-independent) since the former are dimensionless couplings. The effective, matter-induced anomalous scaling dimension $\mathbf{S}_2^\text{(eff)}$ and the term
$\mathbf{A}_{abcd}^\text{(eff)}$ receive contributions from non-vanishing gauge and Yukawa couplings at the fixed point. Their explicit form is given by
\begin{align}
    \mathbf{S}_{2,\;ae}^\text{(eff)} 
    &= 
    -48\pi S_2(\mathbf{\Phi})\alpha_g\delta_{ae} 
    + 8\,\zeta\,\text{Tr}\left(Y_a Y_e\right),\\
    \label{eq:Aabcd}
\mathbf{A}_{abcd}^\text{(eff)}
    &=
    6\,\pi^2\,\alpha_g^2\sum_\text{perms}\{\theta^A,\theta^B\}_{ab}\;\{\theta^A,\theta^B\}_{cd}    
    -2\,\zeta\sum_\text{perms}\text{Tr}\left(Y^a\,Y^{\dagger b}\,Y^c\,Y^{\dagger d}\right)\;.
\end{align}
Here, $S_2(\mathbf{\Phi})$ is the Dynkin-index of the scalar representation, defined by $\text{Tr}(\theta^A\theta^B) = S_2(\mathbf{\Phi})\delta^{AB}$, and $\theta^A$ are the generators of 
the gauge group
in
the scalar representation $\mathbf{\Phi}$.  For group-theoretic details, see, e.g., \cite{Slansky:1981yr}.
Further, $Y^a$ are Yukawa couplings fixed by a Yukawa Lagrangian $\mathcal{L}_{Y} = -\left(Y^a_{ij}\bar\psi_{i}\mathbf{\Phi}_a\psi_j + \text{h.c.}\right)$ and $\zeta=1$ ($\zeta = 1/2$) for Dirac (Weyl) fermions. 
\\

For vanishing gauge couplings $\alpha_g=0$ but non-vanishing (and positive) Yukawa couplings $Y^a_{ij}$, or more generally in a regime of Yukawa dominance, \emph{all} $\mathbf{S}_{2,\;ae}^\text{(eff)}$ are positive, while \emph{all} $\mathbf{A}_{abcd}^\text{(eff)}$ are negative.
Assuming that $A$ and $Y^a_{ij\,\ast}$ are large enough to dominate over all quartic couplings, 
Eq.~\eqref{eq:quarticBeta} simplifies to
\begin{align}
    \label{eq:quarticBeta_limitCase}
  \beta_{\lambda_{abcd}} = 
  \frac{1}{16\pi^2}\mathbf{A}_{abcd}^\text{(eff)}(\alpha_g,\mathbf{Y})
  + A\, \lambda_{abcd}\;.
\end{align}
Since, in the regime of Yukawa dominance, the $\mathbf{A}_{abcd}^\text{(eff)}$ are negative,
there exists a fixed point at which all quartic couplings are positive, i.e., $\lambda_{abcd\,\ast}>0$, and at which they correspond to irrelevant directions. 
In this limit, the fixed point $\lambda_{abcd\,\ast}>0$ generalizes the locally stable fixed-point potential analyzed in the U(1) model of Sec.~\ref{sec:SSB-U1_stab} to more complex gauge groups and scalar representations with multiple distinct quartic invariants.

We caution that it is non-trivial that in a fully coupled system the assumptions of large gravitational contribution $A$ and large fixed-point value of the Yukawa coupling $Y^a_{ij\,\ast}$ can be simultaneously fulfilled.
\\

\subsection{Predictive power for SO(10)}

Now, we are equipped to qualitatively discuss the predictive power of the flow equation \eqref{eq:quarticBeta} for the case of $SO(10)$. 
Starting with the gauge-dominated case with $A<2$, we expect SSB to set in already at the Planck scale, cf.~Sec.~\ref{sec:SSB-U1_stab}.
Since all of the quartic couplings are IR-attractive (for large enough $A$), the direction of the corresponding vacuum expectation value is also fixed, cf.~Sec.~\ref{sec:SSB}. Therefore, this case has the potential to
exclude specific sets of scalar representations. 
For instance, a vacuum expectation value in the $\mathbf{10}$ breaks $SO(10)$ to $SO(9)$, which is not compatible with a viable breaking chain. Hence, the gauge-dominated case might be incompatible with the presence of a scalar transforming in the fundamental representation of the $SO(10)$, unless the interplay with other scalar representations shifts the global minimum away from a vev in the 10-direction.
We tentatively conclude that the scalar $\mathbf{10}$ in $SO(10)$ has to be accompanied by sufficiently large Yukawa couplings which protect it from obtaining a gauge-mediated Planck-scale vacuum expectation value. Indeed, the $\mathbf{10}$ belongs to those representations which allow for Yukawa couplings to the fermionic $\mathbf{16}_F$ representation which contains all Standard-Model fermions.
\\

We now turn to the Yukawa-dominated case,
for which the fixed-point potential 
remains unbroken.
While locally stable, the fixed-point potential is nevertheless non-trivial, i.e., it admits multiple quartic scalar invariants. We demonstrate that such a potential breaks into a particular direction which is predicted by asymptotically safe fixed-point scaling. For SO(10) with Standard Model fermions only, such an example requires (i) all involved scalar representations to admit  
Yukawa couplings to the fermionic $\mathbf{16}_F$ representation containing all Standard-Model fermions to guarantee a stable fixed-point potential around the origin, i.e., positive $m^2$ at the fixed point and (ii) a large enough scalar representation to provide (at least) two non-trivial quartic invariants.
\\

For SO(10), one such representation is the completely antisymmetric 5-index tensor $\mathbf{126}$, cf.~\cite{Mohapatra:1979nn}, which will be denoted by $\phi_{126}$.  
As it is a complex representation, we also include its complex conjugate representation $\overline{\mathbf{126}}$ denoted by $\phi_{126}^*$ such that we can include a mass term $m_{126}^2\,\phi_{126}\phi_{126}^*$. We impose an additional global $U(1)$ symmetry under which $\phi_{126}$ and $\phi_{126}^*$ have opposite charge. This global symmetry  
requires that all scalar invariants contain an equal number of $\phi_{126}$ and $\phi_{126}^*$ fields, in which case the most general (quartic) potential is given by \cite{Buccella:1984ft, Bertolini:2012im}
\begin{align}
    V_{126} \;=\; &
     \lambda_a\,\left[\phi_{126}\phi_{126}^*\right]_{1}^2
    + \lambda_b\,\left[\left[\phi_{126}\phi_{126}\right]_{54}\left[\phi_{126}^*\phi_{126}^*\right]_{54}\right]_{1}
    \notag\\
    &+ \lambda_c\,\left[\left[\phi_{126}\phi_{126}\right]_{1050}\left[\phi_{126}^*\phi_{126}^*\right]_{\overline{1050}}\right]_{1}
    + \lambda_d\,\left[\left[\phi_{126}\phi_{126}\right]_{4125}\left[\phi_{126}^*\phi_{126}^*\right]_{4125}\right]_{1}\;.
\end{align}
Here, $[]_{x}$ denotes contractions in the symmetric tensor product $(\mathbf{126}\times\textbf{126})_\text{symm} = \mathbf{2772} + \mathbf{4125} + \mathbf{1050} + \mathbf{54}$ along the corresponding representation. 
A more explicit form of the contractions is given in App.~A of \cite{Bertolini:2012im}. As is also discussed in \cite{Bertolini:2012im}, a Higgs sector of $\textbf{126}$, $\overline{\textbf{126}}$, and $\textbf{45}$ is sufficient to break $SO(10)$ to the Standard Model and constitutes a ``minimal'' non-supersymmetric SO(10) GUT. Here, we will restrict the discussion to whether the first breaking step of $SO(10)\rightarrow SU(5)$ is realized or not.
It turns out that whenever all $\lambda_i$ are positive, the absolute minimum of the potential  
lies
along the $SU(5)$-invariant direction \cite{Buccella:1984ft}. 
\\
We combine this with the general argument that in a sufficiently Yukawa dominated regime such a fixed point has to always exist and is the most IR-stable (i.e., predictive) of all quartic fixed points, cf.~the discussion around Eq.~\eqref{eq:quarticBeta_limitCase}. This implies that whenever the Yukawa couplings dominate, a breaking of the SO(10) symmetry is predicted to occur in the $SU(5)$-invariant direction via the expectation value of the 126-scalar.
Confirmation of these structural arguments requires to derive the full set of perturbative $\beta$-functions for the quartic scalar potential, including gauge and Yukawa contributions. Given these, they can be supplemented by the non-universal contribution from quantum gravity to then confirm the above statements explicitly. Such an analysis will be provided elsewhere.
\\

\section{Outlook and conclusions}
We have discussed the predictive power of asymptotic safety for gravity-GUT settings under the assumption that the gravitational Reuter fixed point persists under the impact of quantum fluctuations of the matter and gauge fields in the GUT. Coupled to asymptotically safe gravity, GUTs are much more restricted than in a setting without gravity. We expect the wealth of potential breaking chains to reduce to a very small (potentially even vanishing) number of admissible chains, once the constraints on the scalar potential arising from asymptotic safety are taken into account. Specifically, quantum-gravity fluctuations fix the scalar potential - potentially up to the quadratic terms, which might remain free parameters of the setting. 
We have highlighted that therefore the effect of gravitational fluctuations in GUTs is threefold:
\begin{itemize}
\item[i)] All quartic couplings are fully determined at all scales.
\item[ii)] For given mass parameters, this selects the direction of spontaneous symmetry breaking. The scale of spontaneous symmetry breaking is a function of the mass parameter.
\item[iii)] For sufficiently strong gravitational interactions, the mass-parameter itself becomes a predicted quantity, resulting in a prediction of the scale of symmetry breaking.
\end{itemize}

 We stress that this is a significant shift in the predictivity of GUT settings: Without gravity, GUTs suffer from a lack of predictivity, resulting in many viable breaking chains, connected to the freedom to choose the scalar potential arbitrarily.
We expect the converse to be true in an asymptotically safe GUT setting: Once the scalar representations are specified, the potential is fixed completely (potentially up to quadratic terms). This should typically result in a uniquely determined pattern of symmetry breaking\footnote{Potentially, several possible patterns remain, depending on the scale at which symmetry-breaking sets in, i.e., depending on the value of the quadratic terms in the potential.}. We thus expect that most scalar representations can actually be ruled out in this setting since they will not lead to a phenomenologically viable pattern of symmetry breaking. 
As an example of phenomenological interest for point ii) we have highlighted that we expect a scalar in the $\mathbf{126}$ representation of SO(10) with sufficiently large Yukawa coupling to break in the SU(5) direction, which is a possible first step of a breaking chain that leads to the Standard Model. \\

The most interesting possibility is within iii), where quantum gravity could provide a dynamical explanation for the stability of the proton: as we have argued, if the mass-parameters in the potential are predicted, they should come out close to the Planck scale, i.e., $M_{\rm GUT}\lesssim M_{P}$ could be an automatic consequence of asymptotic safety. \\

 In the future, testing the viability of breaking chains from an asymptotically safe vantage point appears to be a worthwhile endeavor. We stress that for such practical purposes, no calculation of gravitational fluctuations is necessary. To explore the consequences for specific GUT breaking chains, the addition of a term $A \,\lambda$ to the beta function of a given quartic coupling $\lambda$ suffices, where $A$ parameterizes the effect of metric fluctuations. Future investigations should involve solutions of Eq.~\eqref{eq:quarticBeta}, together with an investigation of the fixed potential, to set the initial conditions of the RG flow.  \\

In functional RG studies, $A>0$, as required for this scenario, is found in a large part of the microscopic gravitational parameter space. 
We may compare the robustness of the prediction of scalar quartic couplings from asymptotically safe gravity to that of other matter couplings: Whereas the calculations of other (beyond) Standard Model couplings require the gravity contribution to take a specific value, the prediction for Higgs quartic couplings only relies on the \emph{sign} of the gravitational contribution. While the precise value of the gravitational contribution is subject to systematic uncertainties, these do not lead to corresponding uncertainties in the scalar potential, provided that $A$ is large enough \footnote{In the regime of small $A$, the transplanckian values of gauge- and Yukawa couplings play a role, which are themselves set by the interplay with quantum gravity.}. The only additional information that is required is the value of the transition scale. Holding the low-energy value of the Planck mass fixed, a variation in the transition scale follows from changes in the gravitational fixed-point values. Accordingly, the prediction of the scalar quartic couplings does not rely on knowing the precise location of the gravitational fixed point, as long as it falls into that part of the gravitational parameter space where $A>0$ holds, non-minimal couplings are not too large, and where the transition scale is roughly equal to the low-energy Planck mass. 
In order to conclusively determine whether the large number of matter fields in a GUT is indeed compatible with an asymptotically safe gravitational fixed point, extended studies are required.\\

 Ultimately, the combination of two ideas could provide an explanation of the seemingly random gauge group of the Standard Model: Starting from a simple gauge group, such as SO(10) or SU(5), asymptotic safety might potentially uniquely single out a breaking chain that necessarily results in the ${\rm SU}(3)_c\times {\rm SU}(2)_L\times{\rm U}(1)_Y$ of the Standard Model. Moreover, even the scale of the spontaneous breaking of the symmetry might be determined. As we have argued, in this case, the scale should come out close to the Planck scale, providing a dynamical explanation for the stability of the proton. We stress that the high predictive power of the asymptotic-safety paradigm does not provide much ``wiggle room" for fundamental physics, as it dramatically reduces the number of free parameters in a GUT setting, and therefore provides a clear pathway to ruling out a given proposed asymptotically safe GUT setting.

\section*{Acknowledgments}
This work is supported by the Deutsche
  Forschungsgemeinschaft (DFG) under grant no.~Ei-1037/1. This
  research is also supported by the Danish National Research
  Foundation under grant DNRF:90. A.~H.~also acknowledges support by the German Academic Scholarship Foundation.  C.~W.~is supported by the DFG research center "SFB 125 (ISOQUANT)".

\bibliographystyle{JHEP}
\bibliography{Bibliography}  
\end{document}